\documentclass{llncs}

\let\subparagraph\subparagraph{}
\usepackage{booktabs}
\usepackage{multirow}
\usepackage{etoolbox, siunitx}
\usepackage[T1]{fontenc}
\usepackage{pbox}
\usepackage{graphicx}
\usepackage{amsmath}
\usepackage{subfig}
\usepackage{booktabs}
\usepackage{siunitx}
\usepackage{blindtext}
\usepackage[space]{cite}
\usepackage{enumitem}
\usepackage[justification=centering]{caption}
\usepackage{hyperref}
\usepackage{floatrow}
  
\newfloatcommand{capbtabbox}{table}[][\FBwidth]

\newcommand{\specialcell}[2][c]{%
  \begin{tabular}[#1]{@{}c@{}}#2\end{tabular}}

\robustify\bfseries
\begin{document}

\title{Classifying document types to enhance search and recommendations in digital libraries}

\titlerunning{Hamiltonian Mechanics}  
%
\author{Authors}
\authorrunning{authors} 
%
\author{
  Aristotelis Charalampous
  \and
  Petr Knoth
}
\institute{CORE, Knowledge Media institute, The Open University\\
\email{\{aristotelis.charalampous, petr.knoth\}@open.ac.uk},\\
}

\maketitle 

\begin{abstract}
In this paper, we address the problem of classifying documents available from the global network of (open access) repositories according to their type. We show that the metadata provided by repositories enabling us to distinguish research papers, thesis and slides are missing in over $60\%$ of cases. While these metadata describing document types are useful in a variety of scenarios ranging from research analytics to improving search and recommender (SR) systems, this problem has not yet been sufficiently addressed in the context of the repositories infrastructure. We have developed a new approach for classifying document types using supervised machine learning based exclusively on text specific features. We achieve $0.96$ F1-score using the random forest and Adaboost classifiers, which are the best performing models on our data. By analysing the SR system logs of the CORE \cite{DBLP:journals/dlib/KnothZ12} digital library aggregator, we show that users are an order of magnitude more likely to click on research papers and thesis than on slides. This suggests that using document types as a feature for ranking/filtering SR results in digital libraries has the potential to improve user experience. 

\end{abstract}

\vspace{-0.3cm}
\section{Introduction}
\vspace{-0.2cm}

Over the last 15 years, there has been a significant growth in the number of institutional and subject repositories storing research content. However, each repository on its own is of limited use, as the key value of repositories comes from being able to search, recommend and analyse content across this distributed network. While these repositories have been established to store primarily research papers, they contain, in fact, a variety of document types, including theses and slides. Services operating on the content from across this repository network should be able to distinguish between document types based on the supplied metadata. 

However, metadata inconsistencies are making this very difficult. As we show later in the study, \textasciitilde$62\%$ of documents in repositories do not have associated metadata describing the document type. Moreover, when document type is specified, it is typically not done using an interoperable vocabulary. 

Consequently, digital library aggregators like CORE \cite{DBLP:journals/dlib/KnothZ12}, OpenAIRE \cite{rettberg2012openaire} and BASE \cite{summann2006bielefeld} face the challenge of offering seamless SR systems over poor quality metadata supplied by thousands of providers. We hypothesise that by understanding the document type, we can increase user engagement in these services, for example, by means of filtering or re-ranking SR systems results. 

In this paper, we develop a novel and highly scalable system for automatic identification of research papers, slides and theses. By applying this identification system, we analyse the logs of CORE' SR systems to see if we can find evidence of users preferring specific document type(s) over others.

The contributions of the paper are:
\begin{itemize}[label=\textbullet]
\item Presenting a lightweight, supervised classification approach for detecting \textit{Research}, \textit{Slides} and \textit{Thesis}, based on a small yet highly predictive set of features extracted from textual descriptors of (scientific) articles, reaching an F1-score of ${96.2\%}$ with the random forest classifier. 
\item A publicly exposed and annotated dataset\cite{aristotelis_petr_2017} of approximately ${11.5k}$ of documents for the sake of comparison and reproducibility. 

\item Proposing a modified CTR metric, balanced QTCTR, to analyse historical SR systems' logs to evaluate user engagement with the proposed content types in digital library systems, showing our users' inclination towards research and theses over slides.
\end{itemize}

The rest of the paper is organised as follows. Firstly, we discuss related work, followed by the presentation of our current data state. Secondly, we outline our approach and present

results of the classification approach and the analysis of current user engagement using our modified CTR metrics. Finally, we end with a discussion before concluding the paper.

\vspace{-0.3cm}
\section{Related work}
\vspace{-0.2cm}

The library community holds traditionally metadata records as a key enabler for resource discovery. Systems, such as BASE and WorldCat\footnote{https://www.worldcat.org/}, have been almost solely relying on metadata in their search services until today. But as such approach, as opposed to services indexing the content, cannot guarantee metadata validity, completeness and quality, nor can achieve acceptable recall \cite{DBLP:journals/dlib/KnothZ12}, some have started to believe that aggregative digital libraries have failed due to the interoperability issues facing OAI-PMH data providers. In fact, \cite{poynder2016CNI} specifically argues that the fact that BASE and OpenAIRE do not (or cannot) distinguish between document types of the records they harvest makes them ``not as effective as users might assume''.  

While automatic document categorisation using structural and content features has been previously widely studied \cite{sebastiani2002machine, qi2009web, ghosh2008combining}, little work has been done on the issue of document type categorisation in the context of digital libraries until the recent study Caragea et al. \cite{caragea2016document}. They experimented with (1) \textit{bag-of-words}, (2) document \textit{URL tokens} and (3) document \textit{structural features} to classify academic documents into several types. Their set of 43 manually engineered \textit{structural features} have shown significant performance gain over conventional \textit{bag-of-words} models in these highly diverse data collections.

Unlike previous work in standard approaches to text categorisation, summarised in \cite{aphinyanaphongs2014comprehensive}, we use a subset of file and text specific characteristics, selectively gathered from \cite{caragea2016document}. The reduced dimensionality, as a result of the subset's minimal size, allows for scalable integration in ingestion pipelines of SR systems. In addition to the previous work, our study is to our knowledge the first to understand whether the integration of these document type classification systems can lead to more effective user engagement in SR systems. 

\vspace{-0.3cm}
\section{Data - current state}
\vspace{-0.2cm}

CORE is a global service that provides access to millions of (open access) research articles aggregated from thousands of OA repositories and journals at a full text level. CORE offers several services including a search engine, a recommendation system, an API for text-miners and developers as well as some analytical services. As of April 2017, CORE provides access to over $70$ million metadata records and $6$ million full texts  aggregated from $2,461$ data providers. From the available metadata descriptors, a directly available field to categorise records, at a certain extent, is the \textit{dc:subjects} field. While mostly available, currently $92\%$ of cases, only a small minority contain clear descriptions of the document type. More specifically, \textasciitilde$\num[round-precision=1]{30.03062}$ of records are marked as \verb|article|, \textasciitilde$7.3\%$ are marked as \verb|thesis| and $0\%$ as \verb|slides|. This means that we do not have any type document type indication for \textasciitilde${62\%}$ of our data. 
\begin{table}
	\centering
	\vspace{-1\baselineskip}
	\begin{tabular}{lr}
      \toprule
        {Term name} & {Term frequency}
        \\
        \midrule
      	article & \num{0.1365669182}\\
        info:eu-repo/semantics/article & \num{0.0866117754}\\
        journal articles & \num{0.03849475448}\\
        thesis & \num{0.020513418}\\
        info:ulb-repo/semantics/openurl/article & \num{0.0016581145}\\
        info:eu-repo/semantics/doctoralthesis & \num{0.01059937032}\\
        info:eu-repo/semantics/bachelorthesis & \num{0.01010210787}\\
        \midrule
      	\caption{Most popular terms found in the \textbf{dc:subjects} field with >1\% occurrence}
    	\label{term:freq}%
	\end{tabular}
\end{table}

Table \ref{term:freq} lists the top re-occurring terms that are most indicative of the three document types we are interested in. This provides empirical evidence of the poor adoption of interoperable document type descriptors across data providers. Finally, from the \textasciitilde$6$ million full text entries that CORE contains, ${8.5}$ million unique \textit{dc:subjects} field terms are currently recorded (one record can contain multiple subjects fields).

\vspace{-0.3cm}
\section{Approach}
\vspace{-0.2cm}

While one approach to address the problem of poor or missing document type descriptors can be to create guidelines for data providers, we believe this approach is slow, unnecessarily complex and does not scale. Instead, we aim to develop an automated system that infers the document type from the full text. 

The assumptions we make for this study follow several observations on the textual features of documents stored in CORE:

\begin{itemize}[label=\textbullet]
\item \textbf{F1: Number of authors}: The more authors involved in a study, the more likely a document is a research paper as opposed to slides or thesis. 
\item \textbf{F2: Total words}: These were tokenised from the parsed text content using the \textit{nltk}\cite{bird2006nltk} package. Intuitively, the lengthier a document is, in terms of total written words and amount of pages, the more likely it is a thesis.
\item \textbf{F3: Number of pages}: Research papers tend to have a fewer number of pages compared to theses and slides.
\item \textbf{F4: Average words per page}: Calculated as $\frac{\mbox{\#total words}}{\mbox{\#total pages}}$. The fewer words written per page on average, the more likely the document type is \textit{slides}.
\end{itemize}

We extract F2-F4 from their respective \verb|pdf| files with pdfMiner\cite{shinyama2015pdfminer}. F1 is extracted from the supplied metadata. We then apply one of the classifiers, described later in Section \ref{evaluation}, to predict the document type given these features. 

\vspace{-0.3cm}
\section{Experiments}
\vspace{-0.2cm}

\subsection{Data Sample}
\vspace{-0.2cm}
\label{data-sample}

Our first goal was to create a sufficiently large ground truth dataset. Data labelling took place with a rule-based method applied to the CORE dataset. More specifically, we used a set of regular expressions on the {dc:subjects} field and the document's title as follows:

\begin{itemize}[label=\textbullet]
\item Subjects fields for which entries include the keyword ``thesis'' or ``dissertation'' were labelled as \textit{Thesis}.
\item Subjects fields for which entries do \textbf{not} include the keyword ``thesis'' or ``dissertation'' and their title does \textbf{not} include the keyword ``slides'' or ``presentation'' were labelled as \textit{Research}.
\item Subject fields for which entries do \textbf{not} include the keyword ``thesis'' or ``dissertation'' and their title includes the keyword ``slides'' or ``presentation'' were labelled as \textit{Slides}.
\end{itemize}

While this rule-based labelling process produced a sufficiently large number of samples for the \emph{Research} and \emph{Thesis} classes, it has not yielded a satisfactory sample size for the \textit{Slides} class. To address this issue, we have mined \verb|pdf|s and metadata from SlideShare\footnote{https://www.slideshare.net/} using their openly accessible API. 

We wanted the total size of the sample to satisfy two criteria, a confidence level of $95\%$ at a confidence interval of $1\%$. The equation to calculate the necessary size of the data sample is:

\begin{equation}
n=\frac{Z^2\hat{p}(1-\hat{p})}{c^2}
\end{equation}

where, $Z$ is the $Z$ score, $\hat{p}$ is the percentage probability of picking a sample and $c$ is the desired confidence interval. Given a $Z$ score of $1.96$ for a $95\%$ confidence level, a confidence interval of $0.01$ and a sample proportion $p$ of $0.5$ (used as it is the most conservative and will give us the largest sample size calculation), this equation yields \textasciitilde${9.6k}$ samples.

We have gathered these $9.6k$ samples and additionally extended the dataset by $20\%$ to form a validation set, resulting in $11.5k$ samples. To produce a sample with a representative balance of classes, we limited slides to take up to $10\%$ of the final dataset, $55\%$ for research and the remaining $35\%$ for theses entries. We also ensured that all the \verb|pdf|s in the data sample are parsable by pdfminer. 

Finally, we addressed the issue of missing values for feature F1, which SlideShare did not provide in over $97\%$ of cases, by applying multivariate imputations \cite{buuren2011mice}. To improve our knowledge of the feature distributions prior to applying the imputations for the \textit{Slides} class, we relied on extra data from Figshare\footnote{https://figshare.com/}.

\begin{figure}[t]
	\vspace{-1\baselineskip}
    \subfloat[]{{\includegraphics[width=\textwidth/2]{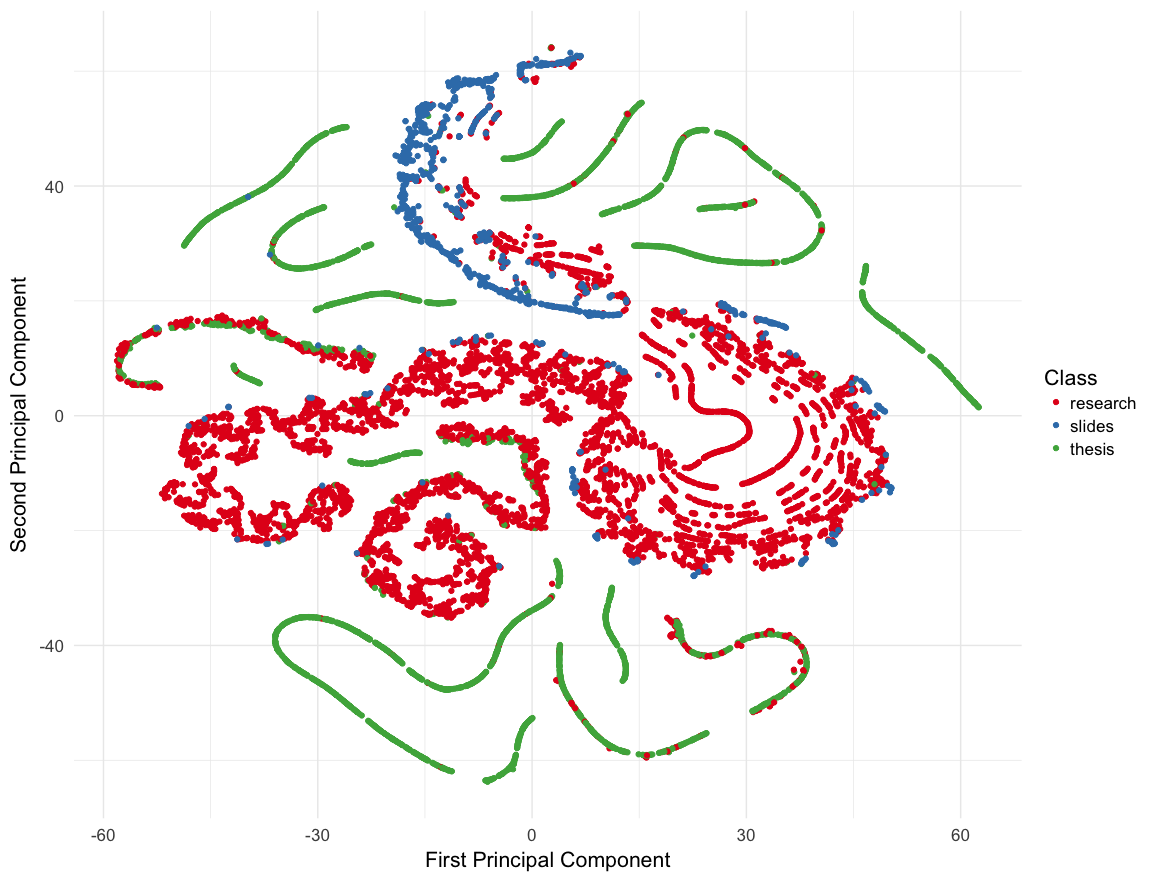} }}%
    \subfloat[]{{\includegraphics[width=\textwidth/2]{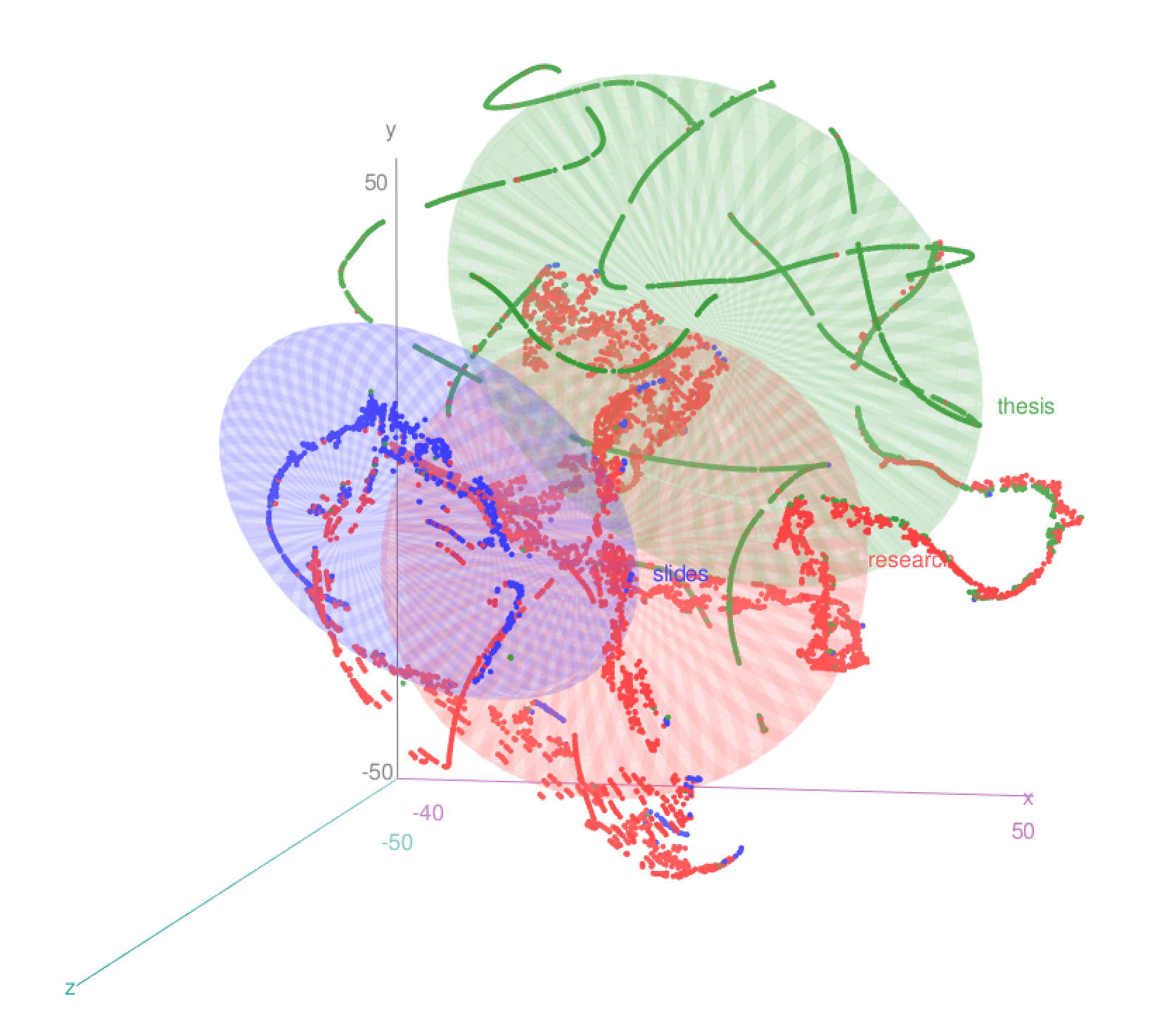} }}%
    \caption{Data variance visualisation using (a) two and (b) three dimensional projections on the corresponding principal components.}%
    \label{tsne:viz}%
\end{figure}

To visualise the dimensionality and data variance in the resulting dataset, we have produced two and three dimensional projections of our data, using techniques introduced by \cite{maaten2008visualizing}. On small datasets ($<\num{100}k$ data points) these do not require much tuning of hyper-parameters and, out of manual inspection from a limited range of hyper-parameters, we decided to use \verb|perplexity| of $30$ and a \verb|theta| of $0.5$. As Figure \ref{tsne:viz} suggests, there is sufficient evidence of data sparsity.

\subsection{Feature Analysis and Model Selection}
\vspace{-0.2cm}
\label{evaluation}

We have experimented with: Random Forest (RF), Gaussian Naive Bayes (GNB), $k$ Nearest Neighbours ($k$NN), Adaboost with Decision trees (Adaboost) and linear kernel Support Vector Machines (SVM). 

We followed a standard 10-fold cross-validation approach to evaluate the models with an extra 20\% of the data left aside for model validation. The class balance discussed was preserved in each fold evaluation by applying stratified splits on both test and validation sets, simulating a representative distribution of categories in the CORE dataset. All features used were compared against their normalised and log-scaled counterparts to check for any possible performance improvements. We have also optimised for a small range of hyper-parameters for each machine learning algorithm using parameter sweeps, recording the best achieved performance for each algorithm class. The evaluation results are presented in Table \ref{train:test:results}.

Two baseline models have been used to assess the improvement brought by the machine learning classifiers. The approaches used are:

\begin{itemize}[label=\textbullet]
\item \textbf{Baseline 1:} Random class assignment with probability weights corresponding to the dataset's class balance. 
\item \textbf{Baseline 2:} A rule-based approach based on statistically drawn thresholds for each feature and class respectively, using the upper $0.975$ and lower $0.025$ quantiles.
\end{itemize}

\begin{figure}[t]
	\vspace{-1\baselineskip}
	\includegraphics[width=\textwidth]{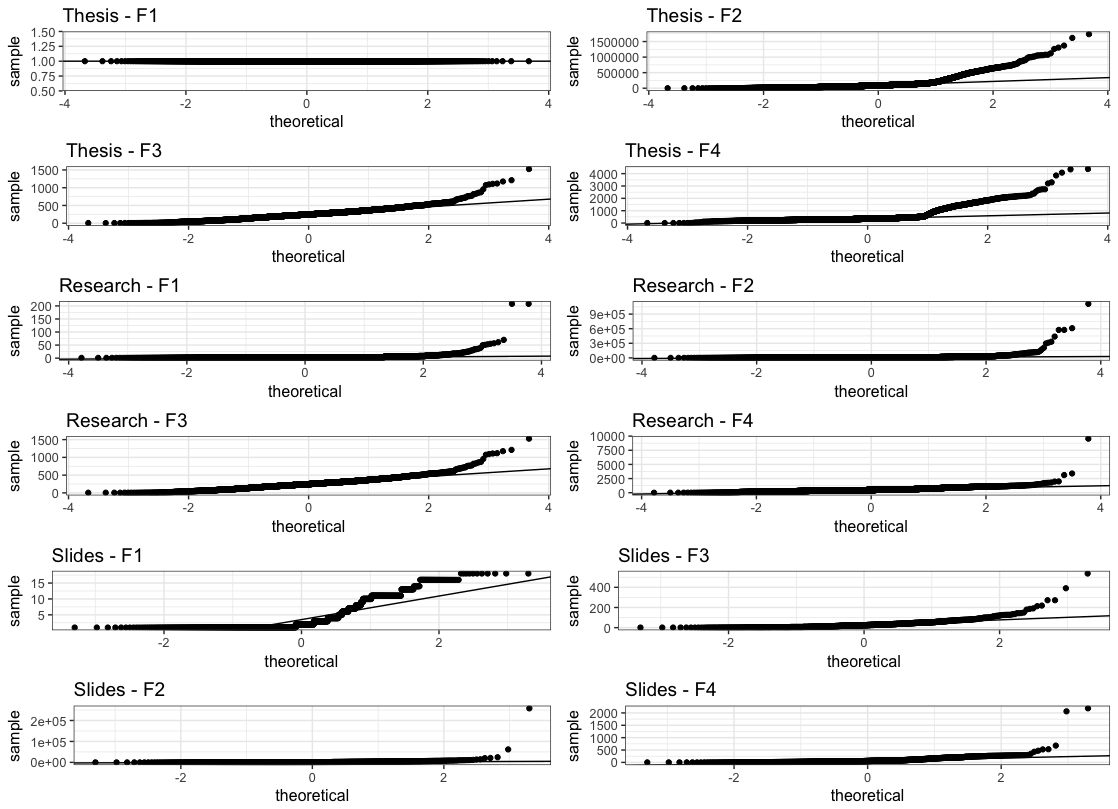}
    \caption{Normal Q-Q Plots split by document type and feature.}
    \label{qq:plots}%
\end{figure}

An analysis was carried out on the assembled dataset to form Baseline 2, based on feature distributions' percentiles. Distributions from the sample dataset largely followed a right skewed normal distribution (Figure \ref{qq:plots}), proving such a model should be a suitable candidate to evaluate against. To avoid overfitting, outliers were removed using Tukey's method \cite{tukey1949comparing}, which was preferred due to its independence on the data distribution, omitting values outside of the range:

\begin{equation}
(Q1 - 1.5*IQR) > Y > (Q3 + 1.5*IQR)
\end{equation}
where, $Y$ is the set of acceptable data points, $Q1$ is the lower quartile, $Q3$ is the upper quartile and $\textit{IQR = Q3 - Q1}$ is the interquartile range.

\begin{table}[H]
	\vspace{-1\baselineskip}
	\begin{center}
    \begin{tabular}{l@{\hspace{1em}}lllll}
      \toprule
      \multirow{2}{*}{Feature} &
        \multicolumn{5}{c}{Document Type}\\
        & {Research} & {} & {Slides} & {} & {Thesis}\\
        \midrule
      F1 & $1 \le x \le 5$ & {} & $1 \le x \le 8$ & {} & ==1\\
      F2 & $\num[round-precision=0]{1226.825} \le x \le \num[round-precision=0]{19151.425}$ & {} & $\num[round-precision=0]{93.6} \le x \le \num[round-precision=0]{7339.8}$ & {} & $\num{15184}\le x \le \num{210720}$\\
      F3 & $3 \le x \le 41$ & {} & $1 \le x \le \num[round-precision=0]{74.575}$ & {} & $47 \le x \le 478$\\
      F4 & $\num[round-precision=0]{208.2297} \le x \le \num[round-precision=0]{926.8950}$ & {} & $\num[round-precision=0]{8.0625} \le x \le \num[round-precision=0]{722.9375}$ & {} & $\num[round-precision=0]{197.7846} \le x \le \num[round-precision=0]{529.9571}$\\
      \midrule
  \end{tabular}
  \caption{Percentile thresholds (upper $0.975$ and lower $0.025$ quantiles) for Baseline 2, following outlier removal.}
  \label{perc:thresh}
  \end{center}
\end{table}

The acquired thresholds for Baseline 2 are listed in Table \ref{perc:thresh}. To assign a particular example a document type $t$, all its features must fall within the boundaries specified. When this method fails, we assign the majority class (\textit{Research}).   

\subsection{Results}
\vspace{-0.2cm}
The evaluation results, presented in Table \ref{train:test:results}, show that all our models outperform the baselines by a large margin. However, baseline 2 demonstrates a perhaps surprisingly good performance on this task. Random forest and Adaboost are the top performers achieving about $0.96$ in F1-score on both the test and validation sets. While we cannot distinguish which model is better at the $95\%$ confidence level and $1\%$ confidence interval, see Section \ref{data-sample}, we decided to productionise random forest due to the model's  simplicity.

\begin{figure}
	\vspace{-1\baselineskip}
	\includegraphics[width=\textwidth]{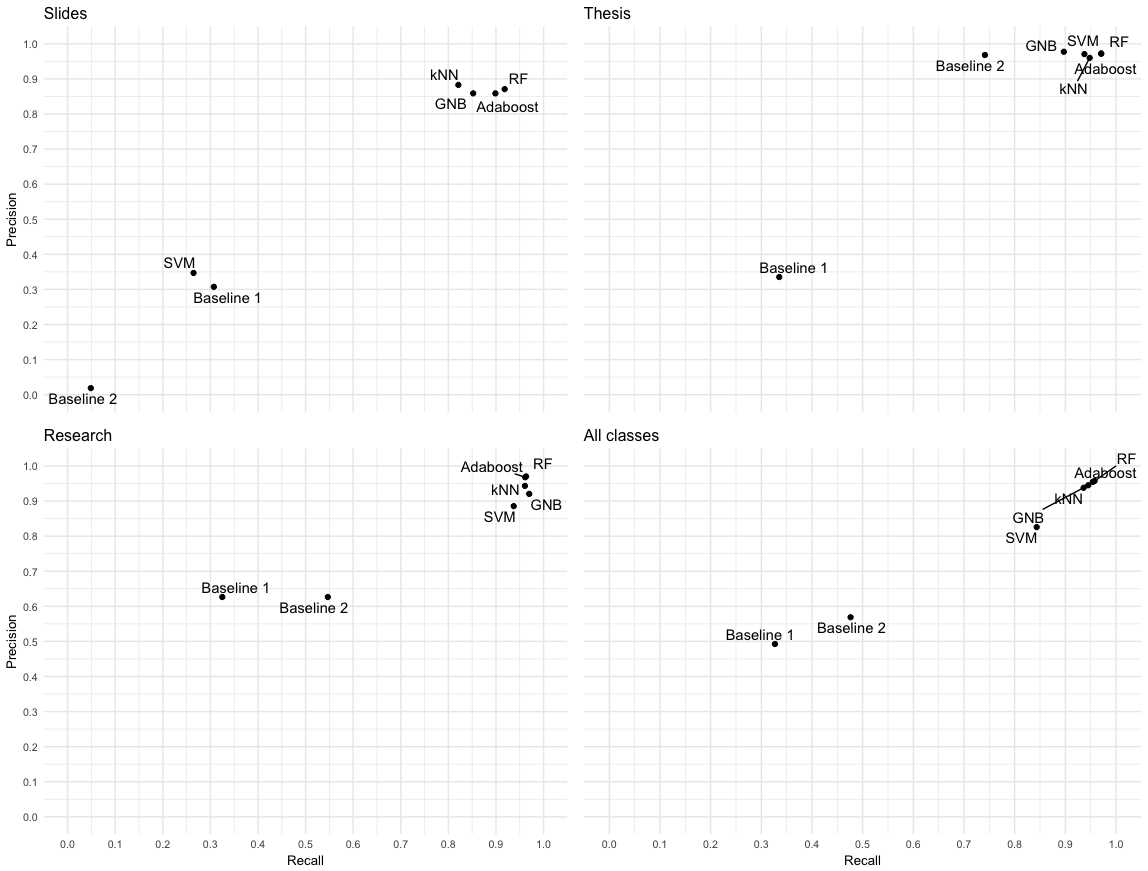}
    \caption{Precision versus Recall for all algorithms on the test set split by class.}
    \label{pvr:viz}%
\end{figure}

Figure \ref{pvr:viz} shows a breakdown of the final precision/recall performances according to the assigned document type. This indicates that a particularly significant improvement of the machine learning models over the baselines is achieved on the \textit{Slides} class. However, as only about $10\%$ of documents in the dataset are slides, the baselines are not so much penalised for these errors in the overall results.

\begin{table}
	\vspace{-1\baselineskip}
	\centering
    \begin{tabular}{l@{\hspace{1em}}lcccccccc}
      \toprule
      \multirow{2}{*}{} &
        \multicolumn{7}{c}{Algorithm}\\
        & {Measure} & {RF} & {GNB} & {kNN} & {Adaboost} & {SVM} & {Baseline 1} & {Baseline 2}\\
        \midrule
      \multirow{3}{*}{\specialcell{Test\\Results}} & Precision & \bfseries \num{0.9623378} & \num{0.9430801} & \num{0.9494912} & \num{0.9580391} & \num{0.8967811} & \num{0.4925689} & \num{0.5687876}\\
      {} & Recall & \bfseries \num{0.9622894} & \num{0.9414282} & \num{0.9497192} & \num{0.9569404} & \num{0.8932870} & \num{0.3269561} & \num{0.4761838}\\ 
      {} & F1-score & \bfseries \num{0.9623000} & \num{0.9416430} & \num{0.9495535} & \num{0.9573248} & \num{0.8695493} & \num{0.3269561} & \num{0.5154115}\\
      \midrule
      \multirow{3}{*}{\specialcell{Validation\\Results}} & Precision & \num{0.9566711} & \num{0.9356236} & \num{0.94531} & \bfseries \num{0.9607432} & \num{0.8434624} & \num{0.5572325} & \num{0.6361684}\\
      {} & Recall & \num{0.9553265} & \num{0.9338488} & \num{0.9454467} & \bfseries \num{0.9604811} & \num{0.8741409} & \num{0.4570447} & \num{0.6564814}\\
      {} & F1-score & \num{0.9557796} & \num{0.9337341} & \num{0.9452725} & \bfseries \num{0.9605947} & \num{0.8310864} & \num{0.4570447} & \num{0.5945136}\\
      \midrule
  \end{tabular}
  \caption{Test and validation set results on weighted evaluation metrics across all algorithms.}
  \label{train:test:results}
\end{table}

To evaluate the importance of individual features, a \textit{post-hoc} analysis was carried out. We fitted the models of our selected algorithms with a \textit{single} feature group at a time. In this scenario, we have recorded high precision performances. Individual feature contributions do not vary widely, except in the case of F4 and the overall performance of the SVM classifier. F1-3 are the most predictive features. We list our findings in Table \ref{indiv:feat:perf}.

\begin{table}[t]
	\vspace{-1\baselineskip}
  \begin{center}
  \begin{tabular}{lSSSSSSSSS}
    \toprule
    \multirow{2}{*}{Features} &
      \multicolumn{9}{c}{Average Weighted F1-score}\\
      & {RF} & {} & {GNB} & {} & {kNN} & {} & {Adaboost} & {} & {SVM}\\
      \midrule
    Only: F2 & \num{0.8825247} & {} & \num{0.7660922} & {} & \num{0.8701925} & {} & \bfseries \num{0.8839478} & {} & \num{0.1868345}\\
    Only: F3 & \num{0.8436393} & {} & \num{0.8412492} & {} & \num{0.8413556} & {} &  \num{0.8424147} & {} & \bfseries \num{0.8441373}\\
    Only: F1 & \bfseries \num{0.8007190} & {} & \num{0.6819066} & {} & \num{0.8007190} & {} & \bfseries \num{0.8007190} & {} & \num{0.6440705}\\
    Only: F4 & \bfseries \num{0.7035729} & {} & \num{0.4744616} & {} & \num{0.6918964} & {} &  \num{0.7017862} & {} & \num{0.3505943}\\
    \bottomrule
    \specialcell{All features\\(RF)} & \bfseries \num{0.9623000} & {} & \num{0.9416430} & {} & \num{0.9495535} & {} & \num{0.9573248} & {} & \num{0.8695493}\\
  \end{tabular}
  \end{center}
  \caption{Classifiers' performance with individual feature groups across all algorithms on the test set in descending order, based on their contribution.}
  \label{indiv:feat:perf}
\end{table}

\vspace{-0.3cm}
\section{Can the model help improve user engagement in SR systems?}
\vspace{-0.2cm}

We applied the random forest model to classify existing content in CORE. Joining the document type information with CORE's SR systems' user logs, enabled us to analyse document type user preferences in CORE's SR systems.\footnote{It should be noted that as CORE provides thumbnails on its SR results pages, users get an idea of the document type prior to accessing it.} We followed the intuition that if we can find that users prefer clicking in SR results on one document type over another, this will provide the argument for using document type information in SR systems to better serve the needs of these users. 

A traditional metric to measure the popularity of a link is the  Click-Through Rate (CTR), measured as:
\begin{equation}
CTR_T = \frac{|Clicks|}{|Impressions|}
\end{equation}

However, we cannot use CTR directly to assess whether people are more likely to click on certain document types than others in the SR system results. This is because we serve, on average, ${\num[round-precision=1]{66.68907}\%}$ \textit{Research}, ${\num[round-precision=1]{27.17887}\%}$ \textit{Thesis} and ${\num[round-precision=1]{6.132052}\%}$ \textit{Slides} impressions across our SR engines. Consequently, the CTR metric would be biased towards the \textit{Slides} class. This is due to the fact that when an action is made on an impression set, the class most represented in the set will benefit from this action on average the least. Put differently, this is accounted to the class imbalance. 

To address this problem, we extend CTR to put \textit{impression equality} into perspective with the following process. We group impressed items in sets $Q$, reflecting the documents served following a query submission (in case of the recommender, the query is a document with respect to which we recommend)\footnote{The number of impressions generated in response to a query can vary across queries. In our case, it can be from zero to ten for search and from zero to five for the recommender.}. We assign to each impression set a type $q_t$ based on the types of document(s) clicked in the results list. In case multiple clicks to distinct document types are made in response to a query, we generate multiple impression sets derived from it, each assigned to one of them. 

We then calculate the \textit{Query Type Click-Through Rate} $(QTCTR)$ as a fraction of the number of queries which resulted in a click to a given document type over the number of all queries:

\begin{equation}
QTCTR=\frac{|Q_T|}{|Q|}
\end{equation}

$QTCTR$ tells us the absolute proportion of queries that result in clicking on a particular document type. We can regularise/normalise $QTCTR$ to reflect the imbalance of impression types, forming the \textit{Regularised Query Type Click-Through Rate} $(RQTCTR)$. We include impression sets with no interaction in this calculation.

\begin{equation}
RQTCTR=\frac{|Q_T|}{|Q|} * \frac{|Impressions_T|}{|Impressions|}
\end{equation}

\begin{table}[t]
	\vspace{-1\baselineskip}
  \begin{center}
  \begin{tabular}{lllcccc||cccccl}
    \toprule
    \multirow{3}{*}{Metric} &
     \multirow{3}{*}{Engine} &
      \multicolumn{8}{c}{Impression set positions}\\
      {} & {} & {} & \multicolumn{4}{c}{Any position} & \multicolumn{3}{c}{Top position}\\
      {} & {} & {} & {} & Research & Slides & Thesis & Research & Slides & Thesis\\
      \midrule
      \multirow{2}{*}{QTCTR} &
      Search & {} & {} & \num[round-precision=5]{0.13685276} & \num[round-precision=5]{0.01878269} & \bfseries \num[round-precision=5]{0.32357637} & \bfseries \num[round-precision=5]{0.038179391} & \num[round-precision=5]{0.003885447} & \num[round-precision=5]{0.018287103}\\
      {} & Recommender & {} & {} & \bfseries \num[round-precision=5]{0.0067483264} & \num[round-precision=5]{0.0007438155} & \num[round-precision=5]{0.0036051059} & \bfseries \num[round-precision=5]{0.0048185582} & \num[round-precision=5]{0.0004610628} & \num[round-precision=5]{0.0020431410}\\
      \midrule
      \multirow{2}{*}{RQTCTR} &
    Search & {} & {} & \num[round-precision=5]{0.081859422} & \num[round-precision=5]{0.001416012} & \bfseries \num[round-precision=5]{0.100611446} & \bfseries \num[round-precision=5]{0.0228372653} & \num[round-precision=5]{0.0002929208} & \num[round-precision=5]{0.0056861132}\\
    {} & Recommender & {} & {} & \bfseries \num[round-precision=5]{0.0048783667} & \num[round-precision=5]{0.0000346020} & \num[round-precision=5]{0.0007927883} & \bfseries \num[round-precision=5]{3.483337e-03} & \num[round-precision=5]{2.144846e-05} & \num[round-precision=5]{4.493011e-04}\\
    \midrule
  \end{tabular}
  \end{center}
  \caption{Modified click-through rate metrics performance on CORE's SR systems.}
  \label{ctr:results}
  \vspace*{-\baselineskip}
\end{table}

The $QTCTR$ and $RQTCTR$ values from the CORE's SR systems, for the three different document types, are presented in Table \ref{ctr:results}. The shows that there is noteworthy difference in preference for \textit{Research} type documents and \textit{Thesis} over \textit{Slides} by an order of one magnitude. This is true for clicks generated on any document in an impression set and when the click was on top positioned document. The $QTCTR$ results also reveal that many people in CORE are looking for theses. We believe this is due to the fact that CORE is one of the few systems (in not the only one) that aggregates theses from thousands of repositories at a full-text level. 

\vspace{-0.3cm}
\section{Scalability analysis}
\vspace{-0.2cm}

There exists a linear relationship between the number of features ($N$) and prediction latency \cite{sklearn:efficiency}, expressed with the complexity of $O(N*M)$, where $M$ are the number of instances. The low number of features and model complexity, with our deployed model having $<10$ trees and $<5$ maximum nodes for each, the latency amounts to slightly over $\num{0.0001}$ seconds per prediction\footnote{This excludes network overhead from the API call and the feature extraction process.}. Due to CORE's continuously ongoing repository harvesting processes, the minimal feature extraction requirements will allow for new additions to be streamlined immediately after their processing, in comparison with the latency associated with the feature extraction process expected from \cite{caragea2016document}. This indicates the high scalability of our approach and applicability across millions of documents.

\vspace{-0.3cm}
\section{Future work}
\vspace{-0.2cm}
In promoting the current solution within CORE's systems, and making it accessible to users worldwide, we aim to:
\begin{itemize}[label=\textbullet]
\item Expose document type classification models as a service, with online model updating, through CORE's public API.
\item Boost \textit{Research} documents in our SR engines and negatively boost \textit{Slides} to aid faster retrieval of preferred content.
\item Evaluate the shift of user engagement as a direct effect of such changes in our services and adjusting our search/recommendation strategies accordingly.
\item Enhance user engagement analysis by cross-validation of our observations here metrics such as the \textit{dwell time}, a metric proven to be less unaffected by position, caption or other form of bias in SR results \cite{kim2014modeling}.
\item Extend the model in further iterations to also discern between sub-types of the \textit{Research} and \textit{Slides} classes, such as theoretical, surveys, use case or seminal research papers as well as slides corresponding to conference papers and lecture/course slides respectively.
\end{itemize}

\section{Conclusions}

We have presented a new scalable method for detecting document types in digital libraries storing scholarly literature achieving $0.96$ F1-score. We have integrated this classification system with the CORE digital library aggregator. This enabled us to analyse the SR system logs of to assess whether users prefer certain document types. Using a our Regularised Query Type Click-Through Rate (RQTCTR) metric, we have confirmed our hypothesis that the document type can contribute in finding a viable solution to improving user engagement.

\section*{Acknowledgements}

This work has been partly funded by the EU OpenMinTeD project under the H2020-EINFRA-2014-2 call, Project ID: 654021. We would also like to acknowledge the support of Jisc for the CORE project.

\bibliographystyle{unsrt}

\bibliography{bibliography.bib}

\end{document}